\documentclass{aa_new}
\usepackage{graphics}
\begin{document}
\title{PLANCK: systematic effects induced by periodic fluctuations of arbitrary shape}

\author{
    Aniello Mennella \inst{1},
    Marco Bersanelli \inst{1,2},
    Carlo Burigana \inst{3},
    Davide Maino \inst{4},
    Nazzareno Mandolesi\inst{3},
    Gianluca Morgante \inst{3,5},
    Giuseppe Stanghellini \inst{3}
} \offprints{A. Mennella}

\institute{
    IFC-CNR, Via Bassini 15, 20133 Milan, Italy \and
    Universit\`a degli Studi di Milano, Via Celoria 16, 20133 Milan, Italy \and
    TESRE-CNR, Via Gobetti 101, Bologna, Italy \and
    Osservatorio Astronomico di Trieste, via G.B.Tiepolo 11, 34131 Trieste, Italy \and
    Jet Propulsion Laboratories, Pasadena, USA
}

\date{Received 2 November 2001  / Accepted 4 January 2002}

\abstract{A fundamental requirement in the new generation of high
resolution Cosmic Microwave Background imaging experiments is a
strict control of systematic errors that must be kept at $\mu$K
level in the final maps. Some of these errors are of celestial
origin, while others will be generated by periodic fluctuations
of the satellite environment. These environment instabilities will
cause fluctuations in the measured signal output thus generating
correlated effects in the reconstructed maps. In this paper we
present an analytical study of the impact of periodic signal
fluctuations on the measured sky maps produced by the {\sc Planck}
survey. In particular we show how it is possible to estimate
analytically the damping factor of the peak-to-peak amplitude of
the fluctuation at the instrument output after the projection in
the final maps. }

\authorrunning{A. Mennella et al.}

\maketitle

\section{Introduction \label{sec:introduction}}

The anisotropy pattern in the Cosmic Microwave Background (CMB)
radiation carries information about fundamental cosmological
properties such as the present expansion rate and average density
of the universe, the amount of dark matter, and the nature of the
seed fluctuations from which all structures in the universe arose
(Vittorio \& Silk \cite{vittorio}, Efstathiou \cite{efstathiou}).
In 1992, the COBE-DMR maps provided the first experimental
evidence of such fluctuations (Smoot et al. \cite{smoot}) at
angular scales $> 7\degr$ at a level $\delta T / T \approx
10^{-5}$ (where $T=2.725\pm 0.002 $~K, Mather et al.
\cite{fixen}). Since then, several balloon-borne and ground-based
experiments have provided CMB anisotropy detection at sub-degree
angular scales, thus improving substantially our knowledge of the
anisotropy power spectrum. Recently, sub-orbital experiments have
been able to map small sky patches with high angular resolution
and sensitivity (e.g. Netterfield et al. \cite{boomerang1}, De
Bernardis et al. \cite{boomerang2}, Lee et al. \cite{MAXIMA1},
Stompor et al. \cite{MAXIMA2}, Halverson et al. \cite{DASI1},
Pryke et al. \cite{DASI2}).

Full-sky CMB maps with high resolution and sensitivity, however,
can be obtained only by space experiments, where the entire sky
is in full view and the environmental conditions are much more
favourable (Danese et al. \cite{danese}). The NASA satellite MAP
(launched in June 2001) will provide the first improved full-sky
maps since COBE, while the ESA satellite {\sc Planck} will map
the CMB with an accuracy set mainly by astrophysical limits.

A key factor for the success of these missions is a strict control
of systematic effects (see, e.g., Bersanelli \& Mandolesi,
\cite{bersanelli}), particularly those that are synchronous with
the spacecraft spin period. These fluctuations, in fact, may be
particularly difficult to remove from the data stream because of
their similarity with the true sky signal, although some data
analysis methods have been developed to deal also with spin
synchronous signals of celestial origin (Delabrouille et al.
\cite{delabrouille}).

In many cases the above effects display a periodic behaviour
leading to a periodic {\em spurious} contribution to the measured
signal. While the transfer function from the fluctuation source
to the measured signal depends on the details of the instrument
and of the effect being considered, the transfer from the
measurement output to the sky map does not depend on the source
but only on the period of the effect and on the scan strategy,
and therefore it can be treated independently.

This work is a study of the impact of periodic signal fluctuations
on the measured sky maps to be produced by the {\sc Planck}
survey. In particular we show how a periodic fluctuation in the
instrument output transfers to the final maps and analyse the
behaviour of the {\em damping factor} versus the oscillation
frequency (Sect.~\ref{sec:from_signal_to_map}). In this section we
also evaluate the possibility to reduce the level of residual
systematic errors by applying destriping algorithms
(Sect.~\ref{sec:destriping}), also considering the combination of
such systematic errors with instrumental noise. The main result of
this analysis is summarised by a general relationship that allows
us to estimate the final peak-to-peak error in a map starting
from a periodic effect of arbitrary shape. In
Sect.~\ref{sec:example} we discuss the application of this
relationship to an example of particular interest for {\sc
Planck}-LFI. Finally, in Sect.~\ref{sec:conclusions} we draw our
main conclusions.

Although the examples presented in this paper are relative to
{\sc Planck}, the concepts at the basis of our treatment are
general, and can be applied to any CMB imaging experiment.

\section{Transfer of periodic oscillations from instrument output to maps\label{sec:from_signal_to_map}}

\subsection{Level of residual fluctuations in scan circle time-scale\label{sec:scan_circle_time_scale}}
The {\sc Planck} measurement strategy is such that the signal
coming from each resolution element in the sky will be measured
approximately 60 times during one scan, thus reducing the effect
of fluctuations occurring on time-scales greater than the spin
period.

In particular, if we average $N$ measurements of the same pixel
taken at times $t_j$ then the residual signal oscillation is given
by:

\begin{equation}
\langle\delta T_{\rm
sky}(t)\rangle_{t_N}=\frac{1}{N}\sum_{j=1}^N\delta T_{\rm
sky}(t_j) \label{eq:av_signal_oscillation}
\end{equation}

\noindent where $t_j = t + (j-1)$ $\tau_{\rm spin}$ ($\tau_{\rm
spin}$ is the spin period which is also the interval between two
consecutive measurements of the same pixel in the sky).

Let us consider a sinusoidal signal oscillation with amplitude
$A_{\rm f}$ and period $\tau_{\rm f}$, i.e. $\delta T_{\rm
sky}(t)=A_{\rm f} \cos (2\pi t/\tau_{\rm f})$.
Eq.~(\ref{eq:av_signal_oscillation}) takes then the form:

\begin{eqnarray}
&&\langle\delta T_{\rm sky}(t)\rangle_{N}= \frac{A_{\rm f}}{N}
\sum_{j=1}^N\cos(2\pi t_j/\tau_{\rm f})=\nonumber \\
&&= \frac{A_{\rm f}}{N} \cos\left[\pi (2\, t + (N-1)\tau_{\rm
spin})/\tau_{\rm f}\right]\frac{\sin(\pi N \tau_{\rm
spin}/\tau_{\rm f})}{\sin(\pi \tau_{\rm spin}/\tau_{\rm f})}
\label{eq:av_signal_sinusoidal_oscillation}
\end{eqnarray}

Eq.~(\ref{eq:av_signal_sinusoidal_oscillation}) gives the
amplitude of the signal oscillation after averaging $N$
measurements of the same pixel starting at time $t$. Because the
term $\cos\left[\pi (2 t + (N-1)\tau_{\rm spin})/\tau_{\rm
f}\right]$ is bounded by $\pm 1$, then the peak-to-peak
fluctuation in the data stream after averaging over $N$ scan
circles is simply given by:

\begin{equation}
\langle\delta T_{\rm sky}^{\rm p-p}\rangle_{N}=
2\left|\frac{A_{\rm f}}{N} \frac{\sin(\pi N \tau_{\rm
spin}/\tau_{\rm f})}{\sin(\pi \tau_{\rm spin}/\tau_{\rm
f})}\right| \label{eq:p2p_av_signal_sinusoidal_oscillation}
\end{equation}

Two special cases can be derived from the above equation: the
case of ``spin synchronous" fluctuations (for which $\tau_{\rm
f}=\tau_{\rm spin}/k$ where $k$ is any integer) and the case of
``spin resonant" fluctuations (for which $\tau_{\rm f} = N\mbox{
}\tau_{\rm spin}/k$, where $k$ is any integer not multiple of
$N$).

Spin synchronous fluctuations are not damped by the measurement
redundancy\footnote{This implication can be easily derived from
Eq.~(\ref{eq:p2p_av_signal_sinusoidal_oscillation}) considering
that $\lim_{x\rightarrow 1}(\sin(N \pi x)/\sin(\pi x))=N$} while
spin resonant fluctuations, instead, are such that for every
pixel the average fluctuation after $N$ consecutive measurements
is zero.

\begin{figure}[here]
\begin{center}
\resizebox{8.5cm}{!}{\includegraphics{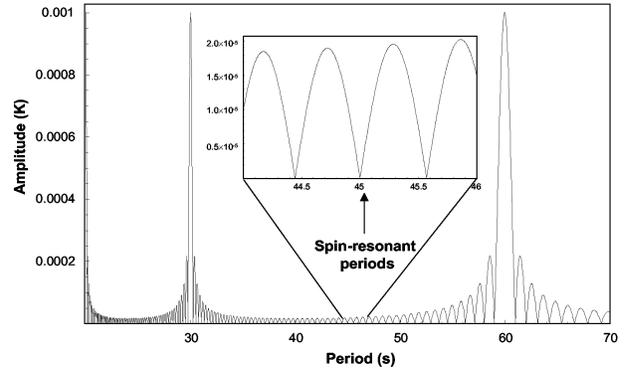}}
\end{center}
\caption{Plot of $\langle\delta T_{\rm sky}^{\rm p-p}\rangle_N$
for for $A_{\rm f}=0.5$~mK and $\tau_{\rm f}$ between 20 s and 70
s. The graph shows the two peaks relative to the spin synchronous
periods of 60 s and 30 s (for which $\langle\delta T_{\rm
sky}^{\rm p-p}\rangle_N=2A_{\rm f}$) and a series of spin
resonant periods (for which $\langle\delta T_{\rm sky}^{\rm
p-p}\rangle_N=0$, see inset).} \label{fig:spin_synch_resonant}
\end{figure}

Let us now consider first oscillations with $\tau_{\rm f}\leq
\tau_{\rm spin}$. In Fig.~\ref{fig:spin_synch_resonant} we show a
plot of $\langle\delta T_{\rm sky}^{\rm p-p}\rangle_N$ given by
Eq.~(\ref{eq:p2p_av_signal_sinusoidal_oscillation}) for values of
$\tau_{\rm f}$ up to 70 s, considering {\sc Planck} nominal values
($\tau_{\rm spin}=60$ s and $N=60$) and assuming $A_{\rm
f}=0.5$~mK.

The picture shows that the residual peak-to-peak fluctuation in a
sky scan circle equals $2A_{\rm f}$ for spin-synchronous periods
and is of the order of $A_{\rm f}/N$ otherwise. The inset in
Fig.~\ref{fig:spin_synch_resonant} highlights that for spin
resonant periods the oscillation is completely cancelled by the
averaging.

\begin{figure}[here]
\begin{center}
\resizebox{8.5cm}{!}{\includegraphics{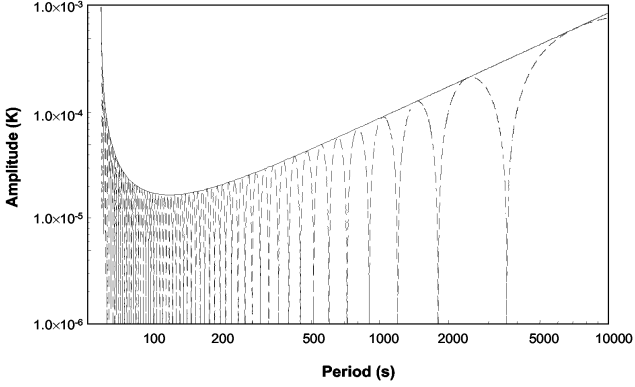}}
\end{center}
\caption{Plot of $\langle\delta T_{\rm sky}^{\rm p-p}\rangle_N$
for for $A_{\rm f}=0.5$~mK and $\tau_{\rm f}$ between 60~s and
10000~s. The graph shows that no more spin synchronous periods
exist after $\tau_{\rm f}=60$~s, while the last spin resonant
period is at $\tau_{\rm f}=3600$~s. The solid line represents the
envelope defined by Eq.~(\ref{eq:envelope}).}
\label{fig:after_60s}
\end{figure}

Fig.~\ref{fig:after_60s} shows the case of oscillations with
$\tau_{\rm f} > \tau_{\rm spin}$. In this case there are no more
spin synchronous periods, but we still have an effect of resonant
periods up to $\tau_{\rm f} = N\times \tau_{\rm spin}$ (which
equals 3600~s in our model). The envelope defining the upper
limit of $\langle\delta T_{\rm sky}^{\rm p-p}\rangle_N$ (solid
line) is given by:

\begin{equation}
\langle\delta T_{\rm sky}^{\rm p-p}\rangle_{N}^{\rm max}=
2\left|\frac{A_{\rm f}}{N} \frac{1}{\sin(\pi \tau_{\rm
spin}/\tau_{\rm f})}\right| \label{eq:envelope}
\end{equation}

Some remarks need to be made concerning the relevance of
spin-resonant and spin-synchronous periods in the context of {\sc
Planck} measurements.

Spin-resonant fluctuations in principle do not contribute to the
overall effect; in reality many non-ideal behaviours (like
instability in the spin period, pointing errors, variations in
the repointing time, etc.) will concur in eliminating such sharp
resonances. The impact of spin-synchronous periods (in particular
$\tau_{\rm f}=60$~s), instead, will be less sensitive to smearing
by scan non idealities. In fact if we consider the main peak in
Fig.~\ref{fig:spin_synch_resonant} we see that it has a width (at
half maximum) of $\sim 1$~s. Because the {\sc Planck} spin rate
will be stable at the level of $\sim 10^{-4}$~rpm/h it follows
that even in the case of a constant drift in the spin rate a
fluctuation which is initially spin-synchronous will maintain
this characteristic for time-scales of the order of one week.

Therefore, in our paper we will adopt a conservative approach by
neglecting only the effect of spin-resonant periods. This implies
that Eq.~(\ref{eq:envelope}) is used to estimate the effect on
the measured data stream of fluctuations with $\tau_{\rm
f}>\tau_{\rm spin}$.

\subsection{Level of residual fluctuations in final maps\label{sec:mission_time_scale}}

The simplest way to estimate the impact of periodic fluctuations
on the final measurements is to project the data stream of
measurements averaged over multiple scans onto a map with a given
pixel size and then evaluate the peak-to-peak amplitude. This
rather simple procedure gives a gross upper limit of the
amplitude of the effect, which is usually greatly reduced by
applying so-called ``destriping" algorithms to the time ordered
data.

In the following we derive a general relationship to estimate the
final peak-to-peak amplitude of a given systematic effect by
knowing its amplitude at the instrument output, its frequency and
the map pixel size. This relationship is obtained analysing the
simple projection of the periodic signal onto the sky map and
then considering the additional damping factor obtained after the
application of our destriping code to the data stream.

\subsubsection{Projection of averaged data streams onto sky maps\label{sec:tod2map}}
The projection of a time ordered array of data onto a map consists
in the association of every data sample to a pixel coordinate in
the sky.

If we consider the {\sc Planck} scanning strategy it is clear
that pixels close to the ecliptic plane will be ``visited" by a
smaller number of scan circles compared to pixels around the
ecliptic poles. Therefore, the  amplitude of a fluctuation
projected onto a sky map will be maximum along the ecliptic,
where each pixel is crossed by $N_{\rm scan}\approx\theta_{\rm
pixel} / \theta_{\rm rep}$ circles, where $\theta_{\rm pixel}$ is
the pixel size and $\theta_{\rm rep}$ indicates the repointing
angle, that in {\sc Planck} current baseline is $\sim 2.5\arcmin$.

A rather simple analysis of the data averaging in pixels around
the ecliptic plane leads to the following formula for the
peak-to-peak amplitude of a fluctuation with amplitude $A_{\rm
f}$ and period $\tau_{\rm f}$ when projected onto a sky map with
pixel size $\theta_{\rm pixel}$:

\begin{equation}
\langle\delta T_{\rm sky}^{\rm p-p}\rangle_{\rm map}=2\frac{A_{\rm
f}}{N\times N_{\rm scan}}\left| \frac{\sin(\pi N\times N_{\rm
scan} \tau_{\rm spin}/\tau_{\rm f})}{\sin(\pi \tau_{\rm
spin}/\tau_{\rm f})}\right|\label{eq:p2p_map}
\end{equation}

From Eq.~(\ref{eq:p2p_map}) it appears that $\langle\delta T_{\rm
sky}^{\rm p-p}\rangle_{\rm map}$ can be zero for every value of
$\tau_{\rm f}$ that satisfies the relationship $\tau_{\rm f} =
N\times N_{\rm scan}\tau_{\rm spin}/k$, with $k$ integer. A set of
these values is what we have called {\em spin resonant} periods,
i.e. those periods which are resonant with the scan circle time
$N\times \tau_{\rm spin}$. Another set of these values
corresponds to periods that are resonant with the repointing time
$\tau_{\rm scan}=N_{\rm scan}\times \tau_{\rm spin}$. Considering
that $N_{\rm scan}$ is not constant all over the map (pixels
close to the ecliptic poles are crossed by a greater number of
scans compared to pixels close to the ecliptic plane) this latter
set of periods does not actually produce a global zero-effect, but
only in a limited number of pixels.

Therefore, with the conservative assumption that spin resonant
periods will not be relevant in practice, we can conclude that
the best estimate of the peak-to-peak amplitude $\langle\delta
T_{\rm sky}^{\rm p-p}\rangle_{\rm map}$ can be well approximated
by the upper limit of Eq.~(\ref{eq:p2p_map}), i.e.:

\begin{eqnarray}
&&\langle\delta T_{\rm sky}^{\rm p-p}\rangle_{\rm map}\approx\langle\delta T_{\rm sky}^{\rm p-p}\rangle_{\rm map}^{\rm max}=\nonumber\\
&& = 2\frac{A_{\rm f}}{N\times N_{\rm scan}}\left|
\frac{1}{\sin(\pi \tau_{\rm spin}/\tau_{\rm
f})}\right|=\frac{2A_{\rm f}}{F_{\rm
map}}\label{eq:p2p_map_envelope}
\end{eqnarray}

\noindent where $F_{\rm map}$ is the damping factor of the
oscillation from the instrument output to the sky map.

We can now compare the damping factors calculated by
Eq.~(\ref{eq:p2p_map_envelope}) with those obtained by projecting
periodic signals with different values of $\tau_{\rm f}$ onto sky
maps. The maps (produced according to the HEALPix hierarchical
structure, G\`orski et al. \cite{healpix}) are relative to the
position of a 30 GHz LFI radiometer and have a pixel size of
13.7\arcmin. All the maps have been produced assuming 10200 hours
observation and the {\sc Planck} baseline scanning strategy,
which implies an average integration time per pixel of $\sim
46.7$~s.

\begin{figure}[here]
\begin{center}
\resizebox{9.cm}{!}{\includegraphics{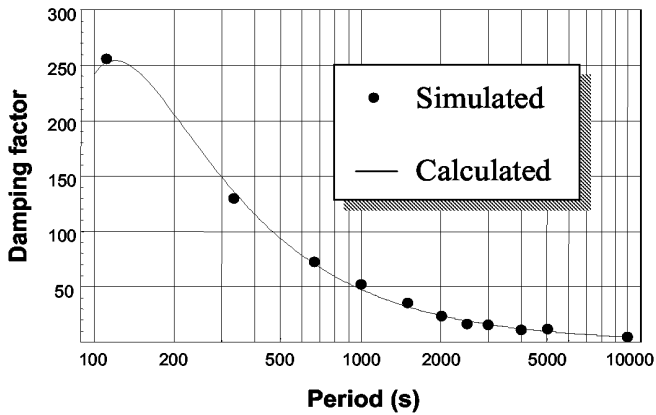}}
\end{center}
\caption{Comparison of the damping factor as a function of
$\tau_{\rm f}$ calculated by Eq.~(\ref{eq:p2p_map_envelope})
(solid line) and obtained from maps (black dots) ($\theta_{\rm
pixel}=13.7\arcmin$, $\theta_{\rm scan}=2.5\arcmin$). }
\label{fig:damping_map}
\end{figure}

Fig.~\ref{fig:damping_map} shows the result of this comparison and
highlights a remarkable agreement between the damping factors
obtained directly from maps and calculated from
Eq.~(\ref{eq:p2p_map_envelope}).

\subsubsection{Removal of periodic systematic effects from Time
Ordered Data \label{sec:destriping}}

In this section we evaluate the residual systematic error on the
maps after the application of a ``destriping" code developed by
the LFI team for reducing the impact of $1/f$ noise fluctuations.

This code generally uses averaged one-hour scan circles (although
it has the capability to work with the full, unaveraged data
streams); because averaging acts like a low-pass filter, only the
very low frequencies remain in the map as a residual effect. The
basic hypothesis is that for each circle the residual effect can
be well approximated by a constant additive level related to the
mean level of the spurious fluctuation during the scan. The code
optimises these levels by minimising the map temperature values in
pixels observed by more than one scan circle during the mission.
Further details of the code implementation can be found in
Burigana et al, \cite{burigana4}.

\begin{figure}[here]
\begin{center}
\resizebox{9.cm}{!}{\includegraphics{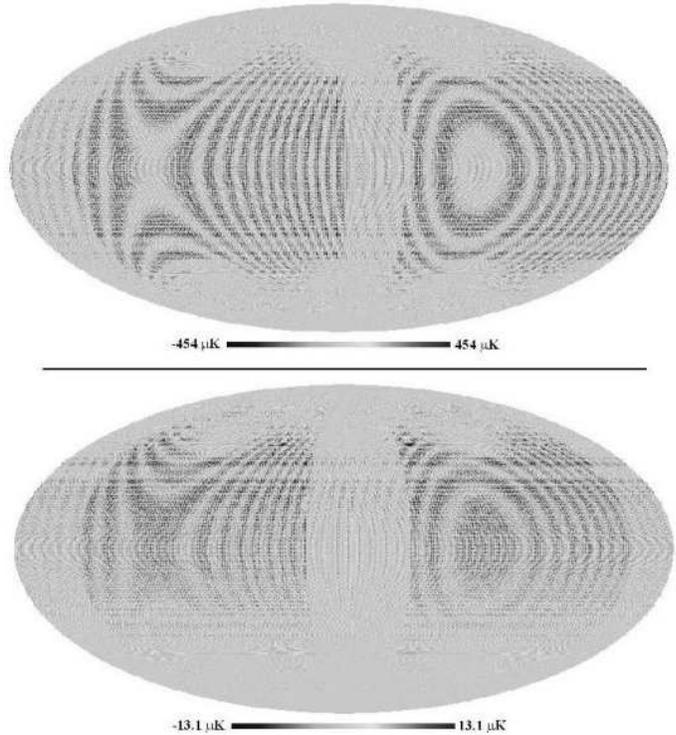}}
\end{center}
\caption{Application of the destriping code to a ``pure" periodic
systematic effect. The upper map is relative to a sinusoidal
fluctuation with a $\tau_{\rm f} =4000$ s and $A_{\rm f}=5$~mK.
After post-processing the time ordered data with our destriping
code the peak-to-peak effect on the map is reduced by a factor
$\sim 34$.} \label{fig:destriping_pure_periodic}
\end{figure}

In Fig.~\ref{fig:destriping_pure_periodic} we show an example of
the application of this code to a pure periodic fluctuation with
a period of 4000 s and an amplitude (at the instrument output) of
1~mK peak-to-peak. A comparison between the two maps shows that
the code is able to reduce the impact of such fluctuation by a
factor of $\sim 34$.

So far we have considered the application of the destriping code
to time ordered data containing only the periodic fluctuation,
i.e. with no noise present. We have also shown that the presence
of white and $1/f$ noise in the data stream does not change the
ability of the code to remove periodic fluctuations. To
verify this we have applied our code to a data stream containing
noise (white and $1/f$) plus a periodic signal and to a second
data stream containing only white and $1/f$ noise.

The noise is typical of LFI 30 GHz radiometers, i.e. white noise
and 1/$f$ noise, with a spectral density of 230~$\mu$K$\,\cdot\,
{\mbox{\rm Hz}}^{-1/2}$ and a knee frequency of 0.1 Hz (rather
conservative estimates of the expected performances 30~GHz LFI
radiometers)\footnote{The reported spectral density is relative
to the noise {\em requirement} of the 30 GHz LFI radiometers and
it has been chosen as a conservative case. The {\em goal} value
is of about 180~$\mu$K~$\,\cdot\,$~Hz$^{-1/2}$ with an expected
knee frequency of 0.05~Hz}. The periodic fluctuation is the same
as in Fig.~\ref{fig:destriping_pure_periodic}.

By taking the difference between the map containing noise plus
the periodic fluctuation and the map containing only noise (both
after destriping) we obtain a map that is virtually
indistinguishable (at the level of $10^{-5}$~$\mu$K) from the
bottom map of Fig.~\ref{fig:destriping_pure_periodic}, which was
obtained by applying the destriping algorithm to the periodic
oscillation without noise. This indicates that the algorithm was
able to recognise and reduce the effect of the periodic
fluctuation even in presence of noise stream with a much larger
amplitude level.

\begin{figure}[here]
\begin{center}
\resizebox{7.cm}{!}{\includegraphics{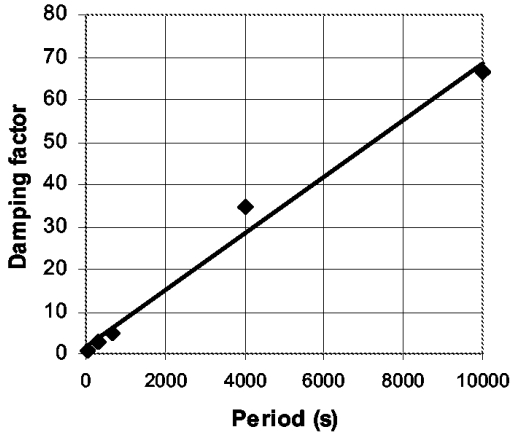}}
\end{center}
\caption{Destriping damping factor for periodic oscillations
versus $\tau_{\rm f}$. Dots indicate the values obtained from
maps while the solid line represents the linear interpolation
$F_{\rm destr}(\tau_{\rm f})=a \tau_{\rm f} + b$ with $a=0.1067$
$\mbox{\rm s}^{-1}$ and $b=1.7992$.} \label{fig:destriping_vs_Tf}
\end{figure}

Our next step has been to evaluate the destriping damping factor
for periodic fluctuations as a function of $\tau_{\rm f}$.
Fig.~\ref{fig:destriping_vs_Tf} shows how this damping factor
increases approximately linearly with $\tau_{\rm f}$ for periods
greater than the spin period\footnote{For spin-synchronous
fluctuations no damping of the effect is obtained from the
application of destriping codes}. It is important to underline
that the exact form of the linear interpolation shown in
Fig.~\ref{fig:destriping_vs_Tf} is dependent on the conditions
under which the maps have been calculated, which are, in the case
discussed here, a sampling rate of 1 sample every $\sim
12$\arcmin \mbox{ } and a map pixel size of 13.7\arcmin \mbox{
}(typical conditions of LFI 30 GHz channels).

If we indicate with $F_{\rm destr}(\tau_{\rm f})=a \tau_{\rm f} +
b$ the destriping damping factor, we can write the following
general form for the total damping factor of the peak-to-peak
amplitude of periodic oscillations from instrument output to
final map:

\begin{equation}
F(\tau_{\rm f}) \geq F_{\rm map}(\tau_{\rm f})\times F_{\rm
destr}(\tau_{\rm f}) \label{eq:total_damping}
\end{equation}

\noindent where $F_{\rm map}$ can be calculated by
Eq.~(\ref{eq:p2p_map_envelope}). Note that
Eq.~(\ref{eq:total_damping}) provides a lower limit of the total
damping factor rather than the exact value. This is a consequence
of our assumptions regarding the beam position and the scanning
strategy which are marginally conservative with respect to the
destriping efficiency.

\begin{figure}[here]
\begin{center}
\resizebox{7.cm}{!}{\includegraphics{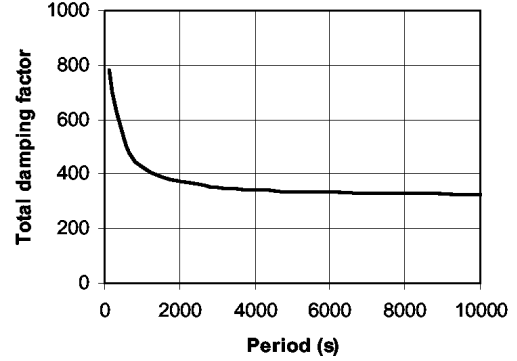}}
\end{center}
\caption{Total damping factor from instrument output to final map
for the peak-to-peak amplitude of periodic oscillations having a
period $\tau_{\rm f} > \tau_{\rm spin}$ considering the destriping
damping as in Fig.~\ref{fig:destriping_vs_Tf}.}
\label{fig:total_damping_vs_Tf}
\end{figure}

In Fig.~\ref{fig:total_damping_vs_Tf} we show a plot of the
function $F(\tau_{\rm f})$ considering the destriping damping
factor relative to the case shown in Fig.
\ref{fig:destriping_vs_Tf}.

\section{Application to a periodic fluctuation of arbitrary shape\label{sec:example}}

In this section we extend our analysis to periodic fluctuations of
general shape and show some examples of its application. Under
very general assumptions we can write the oscillation in Fourier
series as:

\begin{equation}
\delta T_{\rm sky}(t)=\sum_{j=-\infty}^{+\infty}A_j \exp(i 2\pi
\nu_j t) \label{eq:fourier_series}
\end{equation}

\noindent where $\nu_j$ represent the harmonic frequencies. We
now apply the same procedure discussed in the previous sections to
each term of the series in Eq.~(\ref{eq:fourier_series}) in order
to calculate the peak-to-peak amplitude after one-hour averaging
and after the projection of the averaged time data stream onto a
map. Considering that for each harmonic term we have:

\begin{eqnarray}
&&\sum_{k=1}^{N} A_j \exp(i 2\pi \nu_j
(t_0+(k-1)\tau_{\rm spin}))=\nonumber\\
&&= A_j\exp\left[i\pi\left(2 t_0+(N-1)\tau_{\rm
spin}\right)\nu_j\right]\frac{\sin(N\pi \tau_{\rm spin}
\nu_j)}{\sin(\pi \tau_{\rm spin} \nu_j)}
\end{eqnarray}

\noindent then the final peak-to-peak amplitude for a general
signal on the map after destriping can be written as:

\begin{equation}
\langle\delta T_{\rm sky}^{\rm p-p}\rangle_{\rm map}^{\rm
max}=\frac{2}{N\times N_{\rm
scan}}\left|\sum_{j=-\infty}^{+\infty} \frac{A_j/F_{\rm
destr}(\nu_j)}{\sin(\pi \tau_{\rm
spin}\nu_j)}\right|\label{eq:p2p_map_general}
\end{equation}

\begin{figure}[here]
\begin{center}
\resizebox{9.cm}{!}{\includegraphics{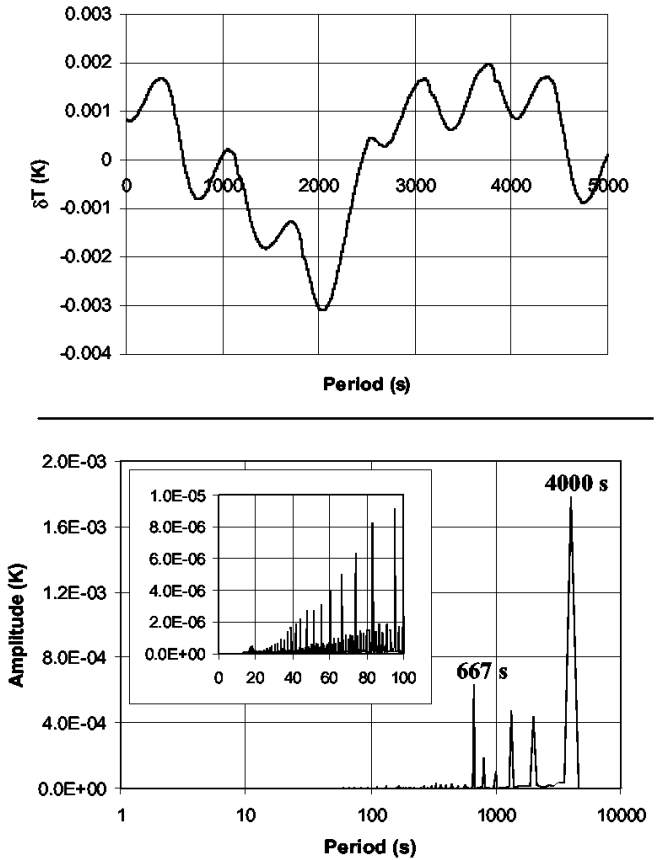}}
\end{center}
\caption{Signal oscillation in time domain (upper graph) and
Fourier space (lower graph) representing a simulation of the
thermally-induced signal instability caused by a Sorption Cooler
working with degraded performances. The inset in the lower graph
is a close-up of the high frequency tail of the fluctuation
spectrum.} \label{fig:general_oscillation}
\end{figure}

Now we show an example of the application of this formalism to the
signal oscillation shown in Fig.~\ref{fig:general_oscillation}
both in time domain (upper graph) and Fourier space (lower
graph)\footnote{Note that in the abscissa axis of the Fourier
transform we indicate the period rather than the frequency}.

This signal oscillation represents an estimate of the
thermally-induced signal instability caused by the {\sc Planck}
Sorption Cooler working with degraded performances. It is worth
underlying that this estimate was derived from
simulations\footnote{Simulations were run using the Sorption
Cooler thermal model implemented with the SINDA simulation
software} which do not represent a quantitative prediction of the
expected Sorption Cooler performances, but were aimed at
understanding qualitatively the impact of compressor
non-idealities on the temperature stability. This oscillation
pattern is used in our paper merely as an example of the
application of our analysis to signal oscillations of complex
shape.

From the figure it is apparent that the fluctuation is dominated
by two main periods (667~s and 4000~s) but it also contains a
high number of harmonics with periods down to about 40~s. The
inset in the lower graph of Fig.~\ref{fig:general_oscillation} is
a close-up of the high frequency tail of the fluctuation spectrum.

\begin{figure}[here]
\begin{center}
\resizebox{9.cm}{!}{\includegraphics{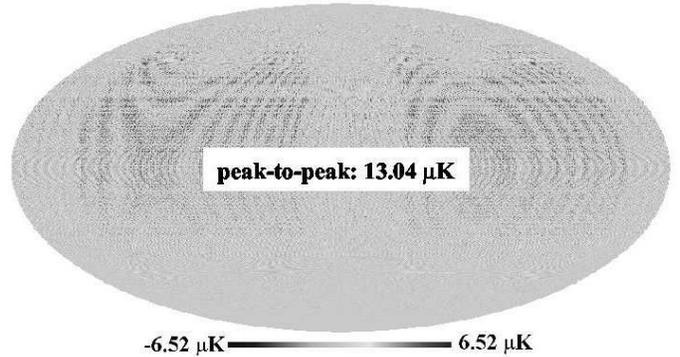}}
\end{center}
\caption{Map (pixel size = 13.7\arcmin) after destriping of the
signal fluctuation plotted in Fig.~\ref{fig:general_oscillation}.}
\label{fig:map_1_bad}
\end{figure}

By applying Eq.~(\ref{eq:p2p_map_general}) to the Fourier
transform of the signal we are able to calculate the expected
peak to peak amplitude of this effect when it is processed by our
destriping code and projected onto a 13.7\arcmin \mbox{ }map. In
the case discussed here the computation yields an expected value
of 13.37~$\mu$K for the final peak-to-peak amplitude, which is in
excellent agreement with the value obtained by the standard
mapping procedure (see Fig.~\ref{fig:map_1_bad}).

Note that in Fig.~\ref{fig:map_1_bad} the morphology of the
residual systematic on the map (which is relative to the signal
oscillation shown in Fig.~\ref{fig:general_oscillation}) is very
similar to the morphology shown in the bottom map of
Fig.~\ref{fig:destriping_pure_periodic} (relative to a ``pure''
periodic signal with $\tau_{\rm f} = 4000$~s). This is the
consequence of the fact that the oscillation in
Fig.~\ref{fig:general_oscillation} is dominated by the 4000~s
wave (as shown in the Fourier transform) which determines the
main visible structure in the map.

The same procedure was also applied to other two signal
fluctuations with very different shapes, which represent
thermally-induced fluctuations for various Sorption Cooler
performance scenarios. The results of the three tests summarised
in Table \ref{tab:comparison_results} show a very good agreement
between the final peak-to-peak amplitudes obtained directly from
maps and calculated with Eq.~(\ref{eq:p2p_map_general}), which
confirms the general validity of our approach.

The values reported in the last two columns of Table
\ref{tab:comparison_results} are relative to 30 GHz LFI maps with
a pixel size $\theta_{\rm pixel}=13.7$\arcmin. Because the optical
beam size will be approximately three times larger than
$\theta_{\rm pixel}$, the actual amplitude of the systematic
effect on the instrument angular scale would be much smaller,
ranging from $\sim 0.5$~$\mu$K to $\sim 5$~$\mu$K for the three
cases reported in Table \ref{tab:comparison_results}.

Although further analysis (which is out of the scope of our paper)
is needed to understand the impact on the recovered amplitude and
polarisation power spectra, our analysis can give rather precise
estimates of the damping factor of the peak-to-peak amplitude of
the fluctuation at the instrument output after the projection in
the final maps.

\begin{table}[here]
\begin{center}
\resizebox{9.cm}{!}{\includegraphics{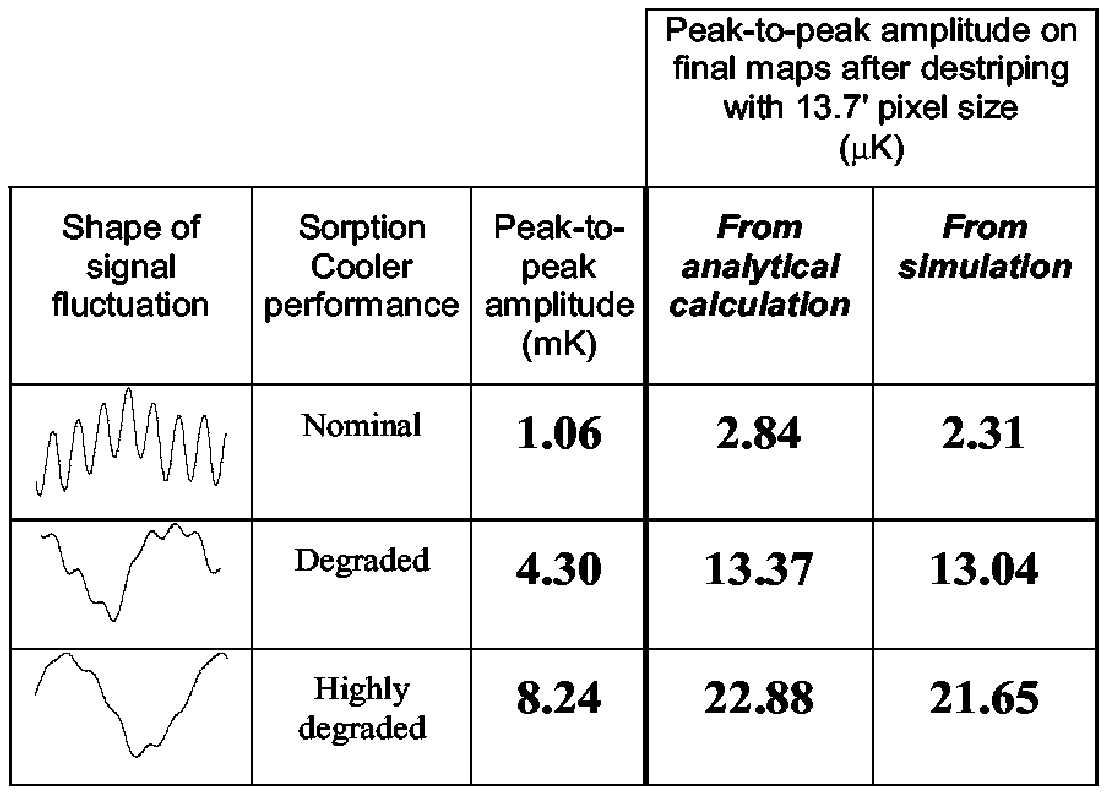}}
\end{center}
\caption{Application of our analytical procedure to signal
fluctuations of different shape and amplitude. These fluctuations
represent the effect of the Sorption Cooler working with
different levels of performance.}\label{tab:comparison_results}
\end{table}

\section{Conclusions \label{sec:conclusions}}
In this paper we have proposed an analytic method to estimate the
effect of periodic signal oscillations on CMB maps to be produced
by sky scanning experiments. When these signal fluctuations are
processed with a destriping algorithm and subsequently projected
onto a sky map, the amplitude is reduced by a factor of several
hundreds; an exception is represented by spin-synchronous
oscillations that, in general, are not damped by the scanning
strategy and can be reduced by dedicated algorithms only in
special cases where we have information about the behaviour of
these signals (e.g. external stray light, Delabrouille et al.,
\cite{delabrouille}).

We have derived an analytical transfer function of the
peak-to-peak amplitude from the signal oscillation at the
instrument output to the map before destriping. The main result
is a general relationship that allows to estimate peak-to-peak
effect of arbitrary periodic fluctuations on maps with arbitrary
pixel size.

The additional damping obtained by applying a destriping
algorithm has been derived as a function of the oscillation
period for maps that are typical of the 30 GHz {\sc Planck}-LFI
channels. The application of this procedure to a signal
fluctuation of complex shape has shown that it is possible to
predict accurately the final peak-to-peak effect.

Although this study has been performed in the context of {\sc
Planck} measurements, the results obtained are generally
applicable to any sky imaging experiment involving redundant
pixel measurements. Further developments of this study will be
aimed at applying these results to predict the impact of
different kinds of periodic systematic effects in {\sc
Planck}-LFI.

\begin{acknowledgements}
Mauro Prina (Jet Propulsion Laboratory, Pasadena, USA)
contributed in simulating the Sorption Cooler temperature
stability characteristics, Roberto Ferretti (LABEN S.p.A.,
Milano, Italy) contributed in simulating the thermal behaviour of
the {\sc Planck}-LFI instrument. Both contributions have been
fundamental to define the signal oscillations at the basis of the
results reported in Table~\ref{tab:comparison_results}.

The HEALPix package use is acknowledged (see HEALPix home page at
http://www.eso.org/science/healpix/). We also wish to thank the
Planck LFI Data Processing Center for the support to the
simulation work.
\end{acknowledgements}

\end{document}